\documentclass[sigconf]{acmart}
\usepackage{mathtools}
\usepackage{balance}

\DeclareMathAlphabet{\mathcal}{OMS}{cmsy}{m}{n}
\AtBeginDocument{%
  }

\copyrightyear{2026}
\acmYear{2026}
\setcopyright{cc}
\setcctype{by}
\acmConference[KDD '26]{Proceedings of the 32nd ACM SIGKDD Conference on Knowledge Discovery and Data Mining V.2}{August 09--13, 2026}{Jeju Island, Republic of Korea}
\acmBooktitle{Proceedings of the 32nd ACM SIGKDD Conference on Knowledge Discovery and Data Mining V.2 (KDD '26), August 09--13, 2026, Jeju Island, Republic of Korea}
\acmDOI{10.1145/3770855.3819014}
\acmISBN{979-8-4007-2259-2/2026/08}

\begin{document}

\title{Reinforcement Learning for Path Integrals in Quantum Statistical Physics}


\author{Timour Ichmoukhamedov}
\affiliation{%
  \institution{University of Antwerp}
  \city{Antwerp}
  \country{Belgium}
}
\affiliation{%
  \institution{New York University}
  \city{New York}
  \state{NY}
  \country{USA}
}
\email{t.ichmoukhamedov@nyu.edu}
\author{Dries Sels}
\affiliation{%
  \institution{Boston University}
  \city{Boston}
  \state{MA}
  \country{USA}}
\affiliation{%
  \institution{Flatiron Institute}
  \city{New York}
  \state{NY}
  \country{USA}
  }
\email{dsels@bu.edu}


\begin{abstract}
Machine learning is rapidly finding its way into the field of computational quantum physics. One of the most popular and widely studied approaches in this direction is to use neural networks to model quantum states (NQS) in the Hamiltonian formulation of quantum mechanics. However, an alternative angle of attack to leverage machine learning in physics is through the path integral formulation, which has so far received far more limited attention. In this paper, we explore how reinforcement learning can be used to compute a class of Euclidean path integrals that yield the thermal density matrix of a quantum system, thereby enabling the computation of the free energy or other thermal expectation values. In particular, we propose a two-step approach with the unique feature that after a variational approximation for a quantity is obtained in a first step, it can then be used to efficiently compute the exact result in a second step. We benchmark this method on several simple systems and then apply it to the quantum rotor chain.
\end{abstract}





\maketitle

\begin{figure*}[t]
  \includegraphics[width=0.95\linewidth]{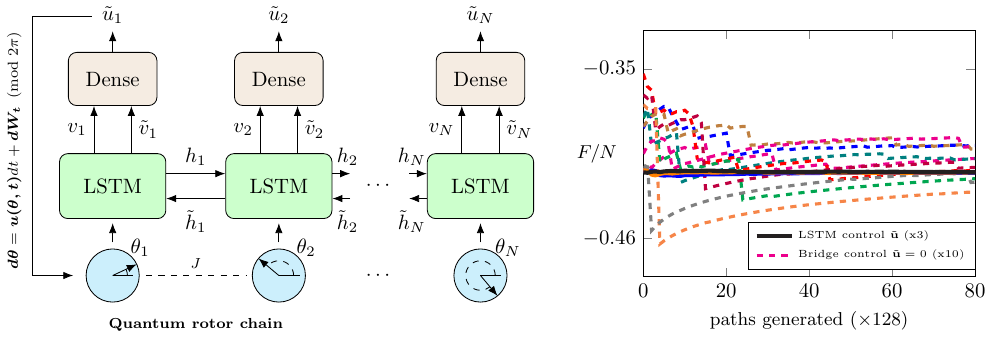}
  \caption{Summary of the central results of this paper. Left panel: we use a shared bidirectional LSTM architecture to train a path integral sampler for the quantum rotor chain (simplified diagram). We show how after training on a fixed number of particles the network can then be readily used on larger systems without having to train again. Right panel (at $J=1$, $\beta=5$) shows several independent direct sampling runs of the free energy with an LSTM trained on $N=9$ but then used for $N=15$. When benchmarked against a simpler bridge control strategy in the absence of a neural net, we observe major improvements in path sampling convergence.}  
  \label{fig:figure1}
\end{figure*}

\section{Introduction}


The prevalent approach in quantum mechanics is to start with a Hamiltonian operator of a given system and find the set of corresponding eigenstates, which can then be used to compute all properties of the quantum system. As such, this is also reflected in the leading paradigm of how machine learning (ML) is currently applied in computational quantum physics, where these states are approximated by neural networks, also known as Neural Quantum States (NQS) \cite{Carleo2017,Torlai2018, Pfau2020,Hibat2020,Sharir2020}. The default and most commonly used NQS formulation is based on the ground state variational principle and has two potential limitations: (1) it can only be used at zero temperature, (2) it is inherently a variational method limited by the neural network after training. While the first point can in principle be addressed with extensions of NQS beyond the ground state \cite{Torlai2018_2,Choo2018, Pfau2024,medvidovic2025adiabatictransportneuralnetwork}, to the best of our knowledge there is no approach within the NQS framework that goes beyond the variational solution except for using it as a guiding function in another downstream method such as Diffusion Monte Carlo~\cite{Inack2018, Ren2023}.

Path integrals provide an alternative description of quantum systems, where instead of finding states, the properties of the system are computed by weighting and summing over all of its trajectories. Although path integrals come in many different forms and are found in quantum dynamics, quantum field theory and classical statistical physics, in this work we are mainly focused on the class of Euclidean imaginary-time path integrals in first quantization, which describe a quantum system in canonical thermal equilibrium. Historically, they have been a valuable complementary tool, and even turned out to be the far more powerful approach in certain settings~\cite{Feynman1955}. While ML has also already been applied to aid with the computation of this class of path integrals, in comparison to NQS far fewer results can be found in the literature in a true quantum physics setting \cite{Che2022,Chen2023,balassa2025}. These approaches have so far not been applied beyond benchmark settings of single particles in one-dimensional potentials and the advantages that ML can offer in this context have not been further explored. 

In this paper, we explore an optimal control approach to computing Euclidean path integrals in physics and show how reinforcement learning (RL) can be used for this purpose. Methodologically, this approach is close in spirit to previous variational RL approaches to compute quantum ground states at zero temperature \cite{barr2020}, with several important modifications. First, we show how this idea needs to be extended towards time-dependent control functions to go beyond the ground state to learn finite-temperature path integral propagators, but also to obtain the partition sum or the free energy of a quantum system. Second, we show that the variational RL solution can be used to kickstart a second direct sampling step that in principle yields exact results for the quantity of interest within the same framework, which provides a unique feature not present in NQS or previous approaches \cite{barr2020}. Importantly, we present applications on a many-body system containing up to 25 particles and also emphasize what type of advantages ML might offer in this context.  

The connection between optimal control and sampling problems is widely known in the literature and goes back several decades \cite{Fleming1981}. More recently, the explicit connection to path integrals has been formalized in a series of papers \cite{Kappen2005_1, Kappen_2005_2, Kappen2016}, and various applications can be found using this connection to solve control problems \cite{Theodorou2010,Theodorou2011,Asmar2022ModelPO, Carius2022}. Optimal control also provides an avenue to solve differential equations with deep learning \cite{Han2018,Nüsken2021} and has been used to sample path integral problems in molecular dynamics \cite{holdijk2023pips}. Recently, stochastic optimal control frameworks similar to the path integral setting \cite{Kappen2016} have also been used in the context of image-generating diffusion models \cite{Domingo2025}. In a closely related setting to ours, the problem of rare event sampling of stochastic processes is solved as an RL problem \cite{Rose2021,Das2021,Gilman2024}. However, to the best of our knowledge, optimal control or RL approaches have so far not been explored to compute Euclidean path integrals of quantum systems at finite temperature, and this will be the central topic of this paper. 

In Sec.~\ref{sec:Methodology}, we introduce the methodology starting with how it can be used to compute individual values of the finite-temperature propagator in Sec.~\ref{subsec:Propagator} and then extend it to the free energy in Sec.~\ref{subsec:Free energy}. As we move through the section, we also present several benchmarking results on simple systems. In Sec.~\ref{sec:Results} we present the main results of this paper in an application to the quantum rotor chain as a model of coupled Josephson junctions \cite{Bradley1984}, which is also a paradigmatic model in the study of quantum phase transitions \cite{sachdev2011quantum}. For this system we compute its free energy and also demonstrate that it is possible to extrapolate the solutions obtained with trained neural networks on smaller system sizes towards larger systems in Sec.~\ref{subsec:extrapolation}. This provides a strong indication that ML approaches, which are precisely known for their interpolation ability, might provide a unique advantage in even larger-scale simulations in the future. In Sec.~\ref{subsec:extrapolation}, we also present the convergence properties of this approach and demonstrate advantages both in terms of path efficiency and computational wall time. Finally, we also show how other thermal expectation values such as the correlation function can be obtained in this approach in Sec.~\ref{subsec:correlation}. 

\section{Methodology}\label{sec:Methodology}

\subsection{Propagator}\label{subsec:Propagator}
The path integral propagator in imaginary time $K(\mathbf{x}_T,T|\mathbf{x}_0,0)$ is defined as the sum over all paths $\mathbf{x}(t) \in \mathbb{R}^n$ on the time domain $\left[0,T\right]$:
\begin{equation}
    K(\mathbf{x}_T,T|\mathbf{x}_0,0) = \int_{\mathbf{x}_0}^{\mathbf{x}_T} \mathcal{D}\mathbf{x} \hspace{1pt} e^{-\mathcal{S}\left[ \mathbf{x}(t) \right]} \hspace{2pt}.
\end{equation}
Assuming units $\hbar=m=1$, we consider action functionals of the form:
\begin{equation}
    \mathcal{S}\left[\mathbf{x}(t)\right] = \int_0^T \left( \frac{1}{2} \left(\frac{d\mathbf{x}}{dt}\right)^2 + V(\mathbf{x}(t)) \right) dt.
\end{equation}
It is not difficult to show that at time $T=\beta$ the propagator is nothing other than the thermal density operator of an interacting many-body system with a general potential $V(\mathbf{x})$ at an inverse temperature $\beta$:
\begin{equation}
    K(\mathbf{x}_T,\beta|\mathbf{x}_0,0) = \langle \mathbf{x}_T| e^{-\beta \hat{H}} | \mathbf{x}_0 \rangle = \rho_{\beta}(\mathbf{x}_T , \mathbf{x}_0 ).
    \label{eq:propagator1}
\end{equation}
It should be noted that we use the unnormalized definition of the thermal density matrix and also that the last equality with the actual $\rho_{\beta}$ only holds for distinguishable particles (e.g. on a lattice). For bosons or fermions, additional permutations need to be taken into account, which is discussed in more detail in Sec.~\ref{subsec:identical}. This provides a clear connection between path integrals and quantum systems in thermal equilibrium. 

The propagator can be numerically computed with the Feynman-Kac theorem \cite{Devreese1996}:
\begin{align} 
K(\mathbf{x}_T,T|\mathbf{x}_0,0) = \mathbb{E}_{ B_0}^{(\mathbf{x}_0)} \left[ e^{-\int_0^T V(\mathbf{x}(t)) dt} \delta\left(\mathbf{x}_T- \mathbf{x}(T) \right) \right],
\label{eq:feynman_kac_propagator}
\end{align}%
where the paths $\mathbf{x}(t)$ are generated with a Wiener process $B_0$  defined by $d\mathbf{x} = d\mathbf{W}_t$, with  $\langle dW_t \rangle= 0$ and $\langle dW^{(i)}_t dW^{(j)}_t \rangle=\delta_{ij} dt$. In the expectation value we highlight that the paths start in a fixed $\mathbf{x}_0$. To account for only selecting the paths ending in the required end point $\mathbf{x}_T$ of the propagator, a delta function $\delta\left(\mathbf{x}_T- \mathbf{x}(T) \right)$ is added, which should for numerical purposes be thought of as for example $e^{-\left(\mathbf{x}_T- \mathbf{x}(T) \right)^2/2\epsilon^2} /(2\pi \epsilon^2)^{d/2}$ for small $\epsilon$. Eq.~\ref{eq:feynman_kac_propagator} provides a concrete numerical prescription for computing a path integral. Random walks need to be generated on a discretized time grid until the expectation value converges, with an exponential suppression for end points away from $\mathbf{x}_T$. Naturally, this sampling scheme will in general have poor convergence properties and the main difficulty is that most paths of $B_0$  will sample regions of the potential that are exponentially suppressed. Moreover, one can expect a non-trivial transition regime from  $t \ll T$ where the optimal sampling region is around the minimum of the potential, towards having to end up in $\mathbf{x}_T$ as $t \rightarrow T$. In what follows we formulate an ML-based approach to learn the optimal sampling scheme for this class of problems. We follow the approach from \cite{Kappen2016} that explored these questions in the context of control problems, but reverse the direction of the derivation and add some modifications.
\begin{figure*}[t]
  \centering
  \includegraphics[width=0.75\linewidth]{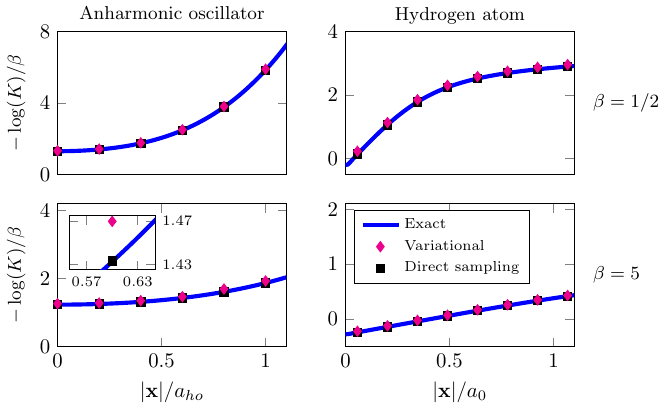}
  \caption{Benchmarking results for the diagonal of the propagator $K(\mathbf{x},\beta|\mathbf{x},0)$ for the anharmonic oscillator $\hat{H}=p^2/2 + x^2/2 + \lambda x^4$ (left at $\lambda=5$) and the hydrogen atom $\hat{H}=\mathbf{p}^2/2 - \frac{1}{|\mathbf{x}|}$ (right) at different temperatures (top row $\beta=1/2$, bottom row $\beta=5$). The inset shows that the variational step (magenta diamonds) indeed provides an upper bound, while the result from the second direct sampling step (black squares) is in agreement with exact diagonalization (blue line).}
  \label{fig:propagator_1}
\end{figure*}
Let us introduce a new stochastic process $B_u$ defined by $d\mathbf{x} = \mathbf{u}(\mathbf{x},t)dt+ d\mathbf{W}_t$ with a control function $\mathbf{u}$ and change the sampling measure in Eq.~\ref{eq:feynman_kac_propagator} to $B_u$. Already anticipating the following steps we can define a cost functional:
\begin{align}
    C_u^{(\mathbf{x}_T)}\left[ \mathbf{x}(t)\right] &= \int_0^T  \left( V(\mathbf{x}(t)) + \frac{1}{2} \mathbf{u}(\mathbf{x}(t),t)^2 \right) dt  \nonumber \\
    &+ \int_0^T \mathbf{u}(\mathbf{x}(t),t) d\mathbf{W}_t - \log( \delta\left(\mathbf{x}_T- \mathbf{x}(T) \right)),
    \label{eq:cost}
\end{align}
where $\log(\delta)$ should be interpreted in a numerical small $\epsilon$ representation of the delta-function. Girsanov's theorem then states that under this new measure Eq.~\ref{eq:feynman_kac_propagator} becomes:
\begin{align} 
K(\mathbf{x}_T,T|\mathbf{x}_0,0) = \mathbb{E}_{u}^{(\mathbf{x}_0)} \left[ e^{-C_u^{(\mathbf{x}_T)}\left[ \mathbf{x}(t)\right]}  \right],
\label{eq:controlled_propagator}
\end{align}
where we use a shorthand notation $\mathbb{E}_{u}$ for paths generated with $B_u$. While Eq.~\ref{eq:controlled_propagator} holds for any function $\mathbf{u}(\mathbf{x},t)$ under the Novikov condition, it turns out that its convergence properties are very much dependent on $\mathbf{u}$. In fact, there exists an optimal control function $\mathbf{u}^*$ such that $B_{u^*}$ is a perfect sampler and Eq.~\ref{eq:controlled_propagator} converges in a single sample \cite{Kappen2016}. Moreover, in principle this function $\mathbf{u}^*$ is only fixed by $\mathbf{x}_T$ and extends fully across $\mathbf{x}_0$ \cite{Kappen2005_1}. It can be shown that the distance to the optimal control function is identified through \cite{Kappen2016}:
\begin{align} 
&\text{KL}\left(  B_u | B_{u^*} \right)   \nonumber \\
& =\log(K(\mathbf{x}_T,T|\mathbf{x}_0,0)) + \mathbb{E}_{u}^{(\mathbf{x}_0)}\left[ C_u^{(\mathbf{x}_T)}\left[ \mathbf{x}(t)\right] \right].
\end{align}
Since the KL-divergence is non-negative, this provides a variational inequality on the propagator that only depends on the control function:
\begin{align} 
-\log(K(\mathbf{x}_T,T|\mathbf{x}_0,0)) \leq \mathbb{E}_{u}^{(\mathbf{x}_0)}\left[ C_u^{(\mathbf{x}_T)}\left[ \mathbf{x}(t)\right] \right].
\label{eq:var_inequality_prop}
\end{align}
\flushbottom

The KL divergence is zero if and only if both distributions are equal, which has two implications. First, for $\mathbf{u}^*$, Eq.~\ref{eq:var_inequality_prop} becomes an equality, and hence the optimal control can in principle be found using a variational approach. Second, this optimal control will also be the perfect sampler for which Eq.~\ref{eq:controlled_propagator} converges in a single sample. Of course, finding the actual optimal control for any sufficiently difficult problem is not feasible, and hence a broader version of this statement is more useful: the better we minimize the variational inequality, the closer the obtained control function will be to the perfect sampler. These observations can now be used to propose a two-step approach to computing the propagator:
\begin{enumerate}
    \item \textbf{Variational}: Propose a variational family of control functions and find one that minimizes $\mathbb{E}_{u}^{(\mathbf{x}_0)}\left[ C_u^{(\mathbf{x}_T)}\left[ \mathbf{x}(t)\right] \right]$. This provides an upper bound on $-\log(K)$ that might already be a sufficient approximation if the expressiveness of the variational family and optimization method are strong enough.
    \item \textbf{Direct sampling}: If an exact result is required, the control function obtained from the variational step, can now be used to sample $\mathbb{E}_{u}^{(\mathbf{x}_0)} \left[ e^{-C_u^{(\mathbf{x}_T)}\left[ \mathbf{x}(t)\right]}  \right]$. This can get an arbitrarily fast convergence in the limit of the variational control function approaching the true optimal control solution.
\end{enumerate}
This implies that path integrals can be viewed as an optimization problem and \textit{exactly solving the optimal control problem is fully equivalent to computing the path integral}. However, finding only an approximate solution is still extremely valuable as it will allow to sample the second step efficiently. 

It is important to note that this optimal control problem can also be written as a differential Hamilton-Jacobi-Bellman equation, which further reduces to the imaginary-time Schr\"{o}dinger equation for this problem \cite{Kappen2005_1,Kappen_2005_2}. Naturally, in high dimensions these equations cannot be solved exactly and the motivation for the present control formulation is to learn approximate or near-optimal control solutions, that could either provide close variational bounds or be leveraged to efficiently sample the exact solution.
\begin{figure}[t]
  \centering
  \includegraphics[width=0.95\linewidth]{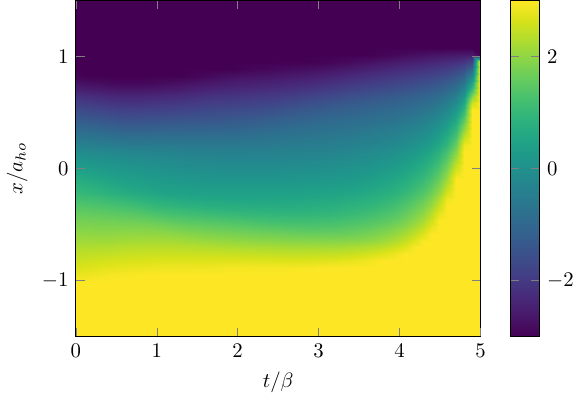}
  \caption{Visualization of the optimized control function $u_\theta$ (including the bridge term) for the anharmonic oscillator at $\lambda=5$, $\beta=5$ and for $x_0=x_T=1$.}
  \label{fig:control_1}
\end{figure}
\subsection{Optimizing the control function with RL}\label{subsec:optimizing}

The optimal control problem can be formulated as a Markov decision process (MDP) by discretizing the time grid on which paths are generated with the Euler-Maruyama method $x_{t_{i+1}}= x_{t_i} + u(x_{t_i}, t_i) dt + dW_t$, where $dW_t$ is a Gaussian increment with mean zero and variance $dt$. The corresponding cost functional in Eq.~\ref{eq:cost} now becomes a sum of (negative) rewards at the corresponding time steps $C_i=V(x_{t_i})dt + \frac{1}{2}u(x_{t_i},t_i)^2dt + u(x_{t_i},t_i)dW_t$ and a terminal cost $(x_T-x(T))^2/2\epsilon^2+d\log(2\pi \epsilon^2)/2$. This problem in MDP form can now be tackled with a broad range of approximate control or RL-like approaches \cite{bertsekas2019reinforcement}. Recently, similar stochastic sampling problems for rare event sampling have been approached with a reinforcement learning approach where both a policy and value function are used to train the sampler \cite{Rose2021,Das2021, Gilman2024}. This type of actor-critic formulation is one of the most common frameworks in modern model-free RL where the value function aids with policy optimization. 

While it is possible to explore a similar approach here, the present problem actually contains more information about both the cost and dynamics than what is available in typical model-free RL settings. Because both the reward and dynamics are differentiable expressions, it is possible to directly estimate the gradients of the cost function with respect to the control function parameters by backpropagating through the entire SDE. This has been done in \cite{barr2020} for quantum ground state computation and is also the approach that will be followed here. One disadvantage is that in large dimensions this method will be memory expensive as it has to keep track of long gradient chains.  We have not run into any memory problems (on a Macbook Air M2 laptop) for the systems studied here and hence proceed with this methodologically simple optimization scheme. For much larger training runs than performed here, various improvements using adjoint gradients \cite{Li2020, Domingo2025} can be expected to achieve better scaling. However, our results in Sec.~\ref{sec:Results} also suggest that with an appropriate parametrization of the control function it might be feasible to use controls trained on smaller systems to sample larger systems, which would partially circumvent this concern. To speed up the training runs we found that without significant loss of accuracy we can use a rougher time grid discretization to train the model while using autograd, and then compute the quantities of interest (including the variational loss) in a validation run on a finer time grid without autograd. 

In our implementation, for a propagator at fixed $\mathbf{x}_T$ we start by writing the control function as:
\begin{equation}
    \mathbf{u}_{\theta}(\mathbf{x},t)= \frac{\mathbf{x}_T-\mathbf{x}}{T-t+\epsilon} + \tilde{\mathbf{u}}_\theta(\mathbf{x},t),
    \label{eq:control_func_1}
\end{equation}
where $\tilde{\mathbf{u}}_\theta$ is parametrized by a neural network. We introduce the driving term $(\mathbf{x}_T-\mathbf{x})/(T-t+\epsilon)$ that we will be referring to as the Brownian bridge term without loss of generality and this term simply ensures that the paths already end up close to the desired end point $\mathbf{x}_T$ when starting with a neural net initiated at zero output. We add a small numerical regularization constant $\epsilon$ which avoids divergence of the control function required for Girsanov's theorem.  Also note that as long as $\mathbf{x}_T$ is fixed, a single optimal control function $\mathbf{u}^*$ considered in notation Eq.~\ref{eq:control_func_1} should generalize across different $\mathbf{x}_0$ (assuming extending training across $\mathbf{x}_0$) without having to condition the control function on it separately. The reason for this is that only $\mathbf{x}_T$ appears in the cost functional, and hence the problem is uniquely determined by a single cost-to-go function $J(\mathbf{x},t)^*$ that can be solved backwards with the Hamilton-Jacobi-Bellman equation back to $\mathbf{x}_0$. In this paper we are mainly interested in the diagonal of the propagator and hence will be training with fixed $\mathbf{x}_0$. For more details we refer to the optimal control treatment of this class of problems \cite{Kappen2005_1,Kappen2016}.  

We find that to discretize the time domain, linearly decreasing time steps $dt$ work better than a uniform grid on $\left[0,T\right]$. This is to be expected since the bridge term is more sensitive towards the end of the path. On this discretized time grid we generate a batch of paths starting in $\mathbf{x}_0$ with the process $B_\theta:d\mathbf{x} = \mathbf{u}_\theta(\mathbf{x},t)dt+ d\mathbf{W}_t$. Since the cost function is fully differentiable, and we keep track of all the gradients, it is straightforward to compute the batch approximation $\approx \nabla_{\theta} \mathbb{E}_{B_\theta}^{(\mathbf{x}_0)} \left[ C_u^{(\mathbf{x}_T)} \left[ x(t)\right] \right]$ by backpropagating through the SDE in any automatic differentiation package (PyTorch in our case).

For more details on the exact numerical implementation, hyperparameters, discretization schemes etc, we refer to our paper repository on Github\footnote{\href{https://github.com/TimourIc/rl4-path-integrals}{https://github.com/TimourIc/rl4-path-integrals}} and Zenodo\footnote{\href{https://doi.org/10.5281/zenodo.20315627}{https://doi.org/10.5281/zenodo.20315627}}. Benchmarking results for simple single particle systems are presented in Fig.~\ref{fig:propagator_1} for the propagator values, and Fig.~\ref{fig:control_1} as an example of the optimized control function. More details on convergence and runtime are explored in a more realistic setting in Sec.~\ref{sec:Results}.

\subsection{Free energy}\label{subsec:Free energy}
\begin{figure}[t]
  \centering
  \hspace*{-1cm}  
  \includegraphics[width=0.95\linewidth]{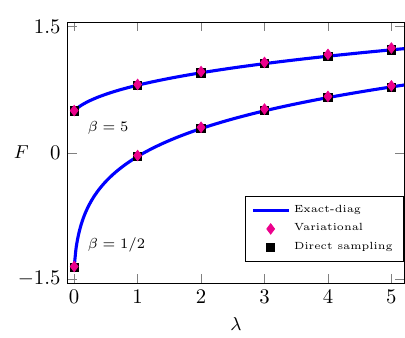}
  \caption{The free energy of the anharmonic oscillator $\hat{H}=p^2/2 + x^2/2 + \lambda x^4$ as a function of the coupling strength $\lambda$, at different temperatures $\beta$. Variational results from Eq.~\ref{eq:frenergy_inequality} (magenta diamonds) are compared with direct sampling Eq.~\ref{eq:dirsample_partition} with the trained $u_\theta(...|\mathbf{z})$, and benchmarked against exact diagonalization (blue line).}
  \label{fig:frenergy_1}
\end{figure}
It is now possible to generalize this approach to the partition function or the free energy. The partition function can be written as a sum over all possible closed paths of a system:
\begin{equation}
    \mathcal{Z} = \int d\mathbf{z} K(\mathbf{z} , \beta| \mathbf{z} , 0) = \mathbb{E}_{\mathbf{z} \sim P(\mathbf{z} )} \left[ \frac{1}{P(\mathbf{z} )} K(\mathbf{z} , \beta| \mathbf{z} , 0) \right],
\end{equation}
where $P$ is some importance sampling distribution to generate the boundary points $\mathbf{z}=\mathbf{x}_0=\mathbf{x}_T$ for the paths. In what follows we will keep $P$ fixed as a simple Gaussian approximation to the diagonal of the propagator and assume that it can be sampled efficiently. By substituting Eq.~\ref{eq:controlled_propagator} we can now write:
\begin{equation}
    \mathcal{Z}  = \mathbb{E}_{\mathbf{z} \sim P(\mathbf{z})} \left[  \frac{1}{P(\mathbf{z} )} \mathbb{E}_{u}^{(\mathbf{z} )} \left( e^{-C_u^{(\mathbf{z})}\left[ \mathbf{x}(t)\right]}  \right) \right].
    \label{eq:dirsample_partition}
\end{equation}
While this equality holds for any control function $u$, the inner expectation value of Eq.~\ref{eq:dirsample_partition} will have varying end points, which are associated with different optimal control functions. Therefore, the appropriate extension of the control function from Eq.~\ref{eq:control_func_1} is to represent an ensemble of control functions across different end points $\mathbf{z}$:
\begin{equation}
    \mathbf{u}_{\theta}(\mathbf{x},t|\mathbf{z} )= \frac{\mathbf{z} -\mathbf{x}}{T-t+\epsilon} + \tilde{\mathbf{u}}_\theta(\mathbf{x},t|\mathbf{z} ).
    \label{eq:control_func_2}
\end{equation}
In practice, $\mathbf{z}$ is of course treated simply as an additional input in a single neural network $\tilde{\mathbf{u}}$, such that it learns the solution to this ensemble of optimal control problems. 
We can now apply the Jensen inequality \cite{Feynman1955} to the inner expectation value to obtain a variational inequality on the free energy:
\begin{equation}
    F \leq  -\frac{1}{\beta} \log\left( \mathbb{E}_{\mathbf{z} \sim P(\mathbf{z} )} \left[  \frac{1}{P(\mathbf{z} )} e^{-\mathbb{E}_{u}^{(\mathbf{z} )}\left[ C_u^{(\mathbf{z} )}\right]} \right] \right).
    \label{eq:frenergy_inequality}
\end{equation}
We can once again use the two-step approach from the previous section and first train $\mathbf{u}_{\theta}(\mathbf{x},t|\mathbf{z})$ by minimizing Eq.~\ref{eq:frenergy_inequality} and obtaining an upper bound on $F$ which might already be a sufficient approximation by itself. This can be achieved by sampling a batch of points $\mathbf{z}$ with respect to $P$, and then sampling paths with respect to $B_u$ from those points using $\mathbf{u}_{\theta}(\mathbf{x},t|\mathbf{z})$. The RHS of Eq.~\ref{eq:frenergy_inequality} is a fully differentiable function and hence we can backpropagate through the SDE just like discussed previously in Sec.~\ref{subsec:optimizing}. While it is indeed true that gradients from some terms can be vanishingly small due to the particular log-exp form of Eq.~\ref{eq:frenergy_inequality}, the importance of those terms can also be expected have a comparably negligible contribution to the free energy. 

To obtain the exact result for $F$ or $\mathcal{Z}$, the trained $\mathbf{u}_{\theta}(\mathbf{x},t|\mathbf{z} )$ can afterwards be used to directly sample Eq.~\ref{eq:dirsample_partition}. The main difference with the previous result is that now there are two sources of sampling variance: (1) the sampling of the closed paths from $P$ to perform the outer trace integral (2) the sampling of paths from $B_u$ to perform the path integral for $K(\mathbf{z}, \beta| \mathbf{z}, 0)$. The optimal control $\mathbf{u}^*$ will now have zero variance associated with the latter path integral, but will not change the variance due to $P$. It should also be possible to make $P$ learnable as part of this scheme and further reduce the variance by exposing it to the sampled values of the propagator. Indeed, during training we observe batches of $(\mathbf{z},K_\theta(\mathbf{z},T|\mathbf{z},0))$ or even $(\mathbf{z},K(\mathbf{z},T|\mathbf{z},0))$, since we can also continuously perform direct sampling with Eq.~\ref{eq:controlled_propagator} on the batches for the latter. Therefore, for a parametrized $P(z)$ (e.g., a Gaussian mixture), it should be possible to iteratively update the parameters on the observed batches. As emphasized before, here we will use a fixed approximation for $P$ for which we already see excellent performance as discussed in Sec.~\ref{sec:Results}. 

As a benchmarking demonstration for this scheme, on Fig.~\ref{fig:frenergy_1} we present both the variational and direct sampling results for the free energy of the anharmonic oscillator. Here, for $P(\mathbf{z} )$ we use a simple Gaussian propagator approximation that is analytically\footnote{Technically, a transcendental equation still has to be solved numerically once.} obtained through the Jensen-Feynman inequality \cite{Feynman1955}.

As these expressions start to get more lengthy, it can be useful to provide an alternative interpretation of the variational inequalities. Starting from the description in Sec.~\ref{subsec:Propagator}, one could define a variational propagator $K_\theta(\mathbf{x}_T,T|\mathbf{x}_0,0)$ that is uniquely induced by the control function as:
\begin{equation}
    u_{\theta}(...|\mathbf{x}_T )  \rightarrow K_\theta(\mathbf{x}_T,T|\mathbf{x}_0,0)= e^{-\mathbb{E}_{u_\theta}^{(\mathbf{x}_0)}\left[ C_{u_\theta}^{(\mathbf{x}_T)}\left[ \mathbf{x}(t)\right] \right]}.
    \label{eq:propagator_induce}
\end{equation}
The inequality in Eq.~\ref{eq:var_inequality_prop} can be written as $-\log(K(\mathbf{x}_T,T|\mathbf{x}_0,0)) \leq - \log(K_\theta(\mathbf{x}_T,T|\mathbf{x}_0,0))$, which holds for a single control function for fixed $\mathbf{x}_T$ across different $\mathbf{x}_0$. This generalizes to the expressions in this section where now the control function is trained across many end-points and hence uniquely induces a complete variational propagator across all initial and end points. In this notation the variational inequality in Eq.~\ref{eq:frenergy_inequality} is nothing else than $F\leq F_\theta$ where $F_\theta$ is the free energy of the propagator induced by $\mathbf{u}_{\theta}(\mathbf{x},t|\mathbf{x}_T )$ through Eq.~\ref{eq:propagator_induce}.

\subsection{Extension to identical particles}\label{subsec:identical}

As emphasized before, the methodology described above applies to either distinguishable particles or lattice degrees of freedom where permutation symmetries do not have to be explicitly accounted for, and the system explored further in Sec.~\ref{sec:Results} will also fall in this category. However, it is illustrative to also briefly discuss how this approach could be extended towards bosons or fermions. For bosons, the many-body path integral propagator can be written as:

\begin{equation}
  K_B(\mathbf{x}_T,T|\mathbf{x}_0,0) = \frac{1}{N!} \sum_P K(\mathbf{x}_T,T | P\mathbf{x}_0, 0),
  \label{eq:boson_prop}
\end{equation}
where $K(\mathbf{x}_T,T | \mathbf{x}_0, 0)$ is the distinguishable particle propagator from the previous sections. The sum is taken over all possible permutations $P$, where if $\mathbf{x}_0=(\mathbf{r}^{(1)}_0, \mathbf{r}^{(2)}_0 ...)$ are the initial points of individual particle positions, $P\mathbf{x}_0$ is a permutation over the particle indices $\mathbf{r}^{(i)}_0$.

\begin{figure}[t]
  \hspace*{-1.2cm}   
  \includegraphics[width=0.95\linewidth]{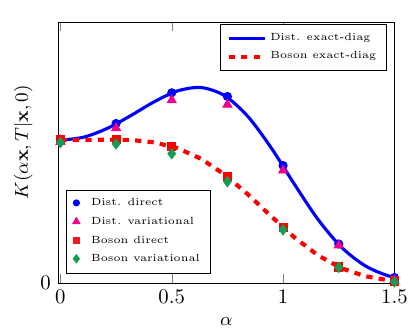}
  \caption{The propagator for $N=3$ one-dimensional bosons at $\beta=1$ in a trapped anharmonic potential with harmonically repulsive interactions $V(\mathbf{x})=\lambda \sum_i x_i^4 -k \sum_{i<j} (x_i-x_j)^2 $ at $\lambda=3$ and $k=5$. The propagator is plotted as a function of a scaled end-point $\mathbf{x}=(0,0.5,1)$ and compared to the non-symmetrized distinguishable (dist.) particle result.}
  \label{fig:boson_propagator}
\end{figure}

One methodologically straightforward way to compute Eq.~\ref{eq:boson_prop} in our approach for tractable $N$ at fixed end point $\mathbf{x}_T$ would be as follows. First, we consider the distinguishable-particle problem and train a fixed control function $\mathbf{u}_\theta(\mathbf{x},t)$ conditioned on the end point $\mathbf{x}_T$ while also sampling different initial points $\mathbf{x}_0$ to provide good learning coverage over the initial conditions. This could, for example, be achieved by permuting $P \mathbf{x}_0$ during training, although this is not a strict requirement as long as a good $\mathbf{u}_\theta$ can be learned. The trained control function $\mathbf{u}_\theta(\mathbf{x},t)$ now induces a variational propagator $ K_\theta(\mathbf{x}_T,T | \mathbf{x}_0, 0)$ and can also be used to directly compute the exact $ K(\mathbf{x}_T,T | \mathbf{x}_0, 0)$ for any initial point $\mathbf{x}_0$. Expression \ref{eq:boson_prop} can therefore be computed by randomly sampling permutations of $P\mathbf{x}_0$, using the trained control function to sample estimators for $K_\theta$ and $K$, and then averaging them over the permutations until convergence. Since all terms in Eq.~\ref{eq:boson_prop} are positive, there is no sign problem and one can expect this method to behave well. Moreover, the average of the variational propagators $K_\theta$ still provides a lower bound (upper bound is for $-\log(K)$) on the actual bosonic propagator $K_B$. We applied this to $N=3$ bosons as an illustration in Fig.~\ref{fig:boson_propagator}. To deal with the rapidly growing number of permutations for larger $N$, one could add another importance sampling step over the permutations that avoids generating long cycles with negligible contribution.  

For fermions, Eq.~\ref{eq:boson_prop} has an alternating sign for odd and even permutations which is the source of the infamous fermionic sign problem \cite{SignProblem} and permutation sampling will not converge in a stable way for larger $N$. A significant amount of previous research effort has been made to formulate path integral identical particle sampling as a diffusion process \cite{Devreese1996,Brosens1995, Luczak1998A,Luczak1998B,Ceperley1991} inside a reduced sector with additional boundary conditions on the edges of the sector. For 1D fermions, this is any sector where the fermions cannot cross each other with absorbing boundary conditions. To learn the propagator for 1D fermions in the present approach, the control function would therefore need to be trained with an additional regularization cost that particles are not allowed to cross. For 3D fermions, the nodes of the sector are not known, and one possible extension would be to train the control function within a fixed-node approximation \cite{Ceperley1991}.

\begin{figure}[t]
  \hspace*{-1.2cm}   
  \includegraphics[width=0.95\linewidth]{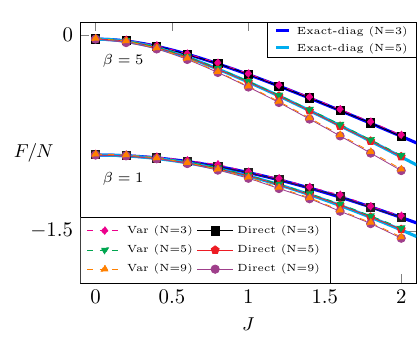}
  \caption{The free energy per particle of the rotor chain at different temperatures $\beta$ and system sizes $N$. For $N=3$ and $N=5$ an exact diagonalization benchmarking result is computed independently (thick lines) which is in excellent agreement with both the variational and direct sampling approaches. For $N=9$ we can no longer obtain the exact diagonalization result but still observe excellent agreement between the variational and direct sampling results.}
  \label{fig:rotor_frenergy_1}
\end{figure}

\section{Results}\label{sec:Results}

\subsection{Quantum rotor chain}

In what follows we will apply this approach to a system where the degrees of freedom exist on a lattice and are therefore distinguishable. One such model is the $O(2)$ (variables on a circle) $d=1$ (one-dimensional) quantum rotor chain that arises in the description of coupled Josephson junctions \cite{Bradley1984}, but also for example describes properties of the Bose-Hubbard model and is more broadly of great interest in the study of quantum phase transitions \cite{sachdev2011quantum}:

\begin{equation}
    \hat{H} = \sum_i  -\frac12 \frac{d^2}{d\theta_i^2} 
    - J \sum_{\mathclap{\langle i,j \rangle}} \cos(\theta_i - \theta_j).
\end{equation}

This Hamiltonian describes a chain of coupled rotors moving on a circle $\theta_i \in [-\pi , \pi [ $ where all neighbors want to align with a coupling strength $J$. The periodic boundary conditions of the rotors require some minor modifications to the approach described in the previous section. First, the stochastic paths need to be generated on this interval mod $2\pi$, which is equivalent to generating paths of all possible winding numbers on the circle. Second, the control function should be made periodic on this interval which means that the Brownian bridge term also acts mod $2\pi$ and in the neural network term of the control this will be enacted by encoding $\sin(\theta), \cos(\theta)$ as an input. For the boundary distribution $P(\mathbf{z})$ we sample the angles from a simple phenomenological proposal ansatz of autoregressive Gaussians (on a circle) relative to the previous angle, with the variance proportional to $\sigma^2=\coth(\beta\sqrt{J}/2)/(2\sqrt{J})$, which is a rough estimate of the expected thermal length scale in the system. 

In the previous benchmarking results shown throughout Sec.~\ref{sec:Methodology}, simple MLPs were used to parametrize $\tilde{u}_\theta$, but for this system we switch to an LSTM autoregressive architecture for which a simplified visualization is depicted in Fig.~\ref{fig:figure1}. The network uses a shared LSTM cell that runs from the left to right of the rotor chain starting with no hidden state on the left, and then the same cell runs from right to left in reverse, now starting with no hidden state on the right. The inputs to the LSTM cell at each particle position $i$ are $\sin(\theta_i),\cos(\theta_i), \sin(z_i), \cos(z_i), t$, with $z_i$ being the sampled boundary point as discussed in Sec.~\ref{subsec:Free energy}. In the actual implementation we stack the LSTM twice and the final outputs are concatenated and passed through a dense layer to obtain the controls. The important difference with how autoregressive architectures might be more commonly used, is that here the hidden states do not run across time, but rather across particles at fixed time. This means that the architecture is in principle indifferent to the number of particles $N$ in the system. The idea for this architecture is inspired by how RNNs can be used to autoregressively sample spin systems across a spin lattice \cite{Hibat2020,Moss2025}. We can now compute the free energy per particle of this system as shown in Fig.~\ref{fig:rotor_frenergy_1} and confirm its accuracy on $N=3$ and $N=5$ with an exact diagonalization benchmarking result. Note that on this figure, a separate network is trained for each of the data points shown on the figure. 

\begin{figure}[t]
  \hspace*{-1.2cm}  
  \includegraphics[width=0.95\linewidth]{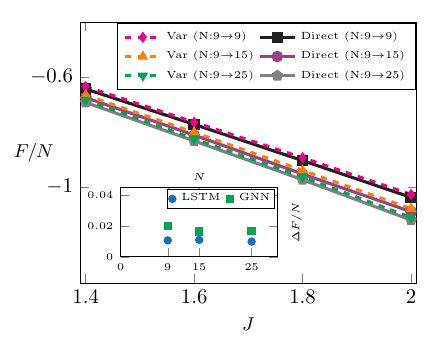}
  \caption{Extrapolation sampling results for the rotor chain free energy ($\beta=5$) where a network is trained for system size $N=9$ and then used for $N=15$ and $N=25$ without training again. Both the variational result and direct sampling results show a decrease in $F/N$ towards larger system sizes and respectively remain in close agreement. The inset shows the variational gap $\Delta F$ between the variational and direct sampling results at $J=2$ as a function of N. For reference, a GNN (inset only) with 3 message passing rounds is also trained and exhibits qualitatively similar extrapolation ability.}
  \label{fig:rotor_chain_extrapolation}
\end{figure}

\subsection{Extrapolation sampling}\label{subsec:extrapolation}

\begin{figure}[t]
  \hspace*{-1.3cm}  
  \includegraphics[width=0.95\linewidth]{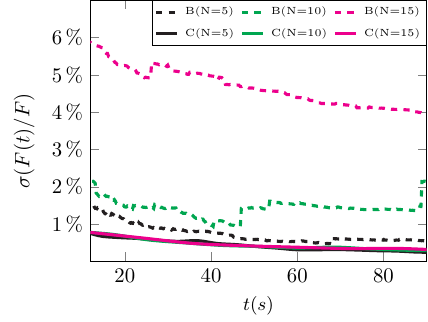}
  \caption{Standard deviation of the rotor chain $F(t)/F$ over 20 independent runs ($J=1$ and $\beta=5$) as a function of the average elapsed wall time per run for direct sampling of the free energy. This represents the relative size of the error bar of a single sampling run at time $t$. The results are computed in the extrapolation sampling setting where the model has only been trained on $N=5$ and compared between $\tilde{u}$-controlled runs (C) and bridge runs $\tilde{u}=0$ (B).}
  \label{fig:rotor_walltime}
\end{figure}

The fact that the proposed architecture does not explicitly depend on the number of particles allows us to explore an avenue that might be particularly well-suited for an ML approach. The autoregressive form of the LSTM means that it can be used to continue generating controls for any $N$ beyond the number it was trained on, and opens up the possibility of using it on larger system sizes without training again, something we will refer to as \textit{extrapolation sampling}. An example of this is shown on Fig.~\ref{fig:rotor_chain_extrapolation} with an extrapolation from $N=9$ to $N=15$ and $N=25$. For this result it is important to remember that the direct sampling approach in Eq.~\ref{eq:dirsample_partition} is valid for any control function $\mathbf{u}$, and hence using the network trained on $N=9$ for the direct sampling of $N=15$ is allowed and should still give exact results. The more important quantity in this context is the variational free energy in Eq.~\ref{eq:frenergy_inequality}, and its closeness to the direct sampling result tells us how close that particular control function is to the optimal control for that system. As can be seen on Fig.~\ref{fig:rotor_chain_extrapolation}, the variational result is comparably equally close when applied to the larger systems and hence provides strong support for the idea of extrapolation sampling. This opens a potentially alternative avenue to perform finite-size scaling extrapolations that are common in Monte Carlo \cite{Binder1984}.

In this context we can now also look at the computational advantage provided by training a neural network for the direct sampling of Eq.~\ref{eq:dirsample_partition}. Because the prospect of extrapolation sampling is to use a trained network across multiple $N$ without training again, in what follows we will not take the training time into account. The comparison is performed against the simple $\tilde{u}=0$ case with only the Brownian bridge, which is still significantly better than the bare Feynman-Kac from Eq.~\ref{eq:feynman_kac_propagator} where the paths are not being driven at all. There are two ways to explore convergence: (1) in terms of number of paths required (absolute metric) and (2) in terms of runtime (setup and implementation dependent). The comparison in terms of path efficiency is shown on the right panel of Fig.~\ref{fig:figure1} for the $N=9 \rightarrow 15$ case where several direct sampling runs on the free energy are compared. It is clear that in terms of path efficiency, having a trained control network, even when doing extrapolation sampling, yields major improvements and converges almost instantly in comparison to the bridge.

However, since using a neural network to generate the paths is also far more costly it is important to confirm that this is also reflected in improvements in actual wall time. This is of course very much dependent on the specific implementation, and here we present the comparison as obtained on a Macbook Air M2 (24 GB RAM) on CPU without any explicit parallelization or other tricks. The results are presented on Fig.~\ref{fig:rotor_walltime} and show the standard deviation of extrapolation sampling from $N=5$ to $N=10$ and $N=15$ over 20 independent runs relative to the actual converged value of $F$. Here we can see a more complete picture. Around $N=5$, sampling with the neural net is only slightly better than the bridge in terms of wall time, but as the system size increases the bridge rapidly worsens while controlled sampling depends only weakly on the number of particles.

\subsection{Correlation function}\label{subsec:correlation}
 
This approach can also be used to compute other thermal expectation values that depend on the propagator. The $O(2)$ and $d=1$ quantum rotor chain introduced in the previous section has a KT-phase transition at $T=0$ \cite{sachdev2011quantum} which can be observed in the correlation length of the system. The equal-time correlation function of the quantum rotor chain can be computed in a direct sampling step
\begin{equation}
  C_{ij}^{(t)} =  \frac{ \mathbb{E}_{\boldsymbol{\theta} \sim P(\boldsymbol{\theta})}  \left[ \frac{1}{P(\boldsymbol{\theta})} \mathbb{E}_{u}^{(\boldsymbol{\theta})}\left[ \cos(\theta_i(t)-\theta_j(t))  e^{ -C_u^{(\boldsymbol{\theta})}[\boldsymbol{\theta}(t)]} \right] \right] } { \mathbb{E}_{\boldsymbol{\theta} \sim P(\boldsymbol{\theta})}  \left[ \frac{1}{P(\boldsymbol{\theta})} \mathbb{E}_{u}^{(\boldsymbol{\theta})}\left[    e^{ -C_u^{(\boldsymbol{\theta})}[\boldsymbol{\theta}(t)]} \right] \right]}
  \label{eq:correlation_func_2}
\end{equation}
which can then be averaged over all times $\langle \cos(\theta_i - \theta_j) \rangle= C_{ij}=\int C_{ij}^{(t)} dt / T $. 

\begin{figure}[t]
  \hspace*{-0.5cm}  
  \includegraphics[width=0.95\linewidth]{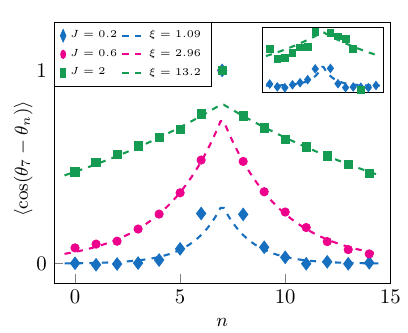}
  \caption{Angular correlation function relative to the central site for an $N=15$ rotor chain ($\beta=5$) computed with Eq.~\ref{eq:correlation_func_2}. The dashed lines represent an $Ae^{-|i-j|/\xi}$ fit for $j\neq i$ with the values for $\xi$ displayed in the legend. Results were computed on 37120(x15) generated paths, and the inset shows the correlation function ($J=0.6$ excluded for visibility) for the same number of paths using the bridge with $\tilde{u}=0$.}
  \label{fig:rotor_correlation}
\end{figure}

Here, we will not be taking the $T=0$ limit to probe the actual KT-transition, but on Fig.~\ref{fig:rotor_correlation} we show that finite-temperature correlation functions can be readily computed using control functions that were trained on the free energy. In particular, Fig.~\ref{fig:rotor_correlation} was also computed in an extrapolation sampling setting where the model trained on $N=9$ was used. We can see that for an equal number of paths the network-controlled computation has nearly fully converged while the correlation values produced by the bridge on the inset have still not even fully converged inside the axis limits at strong coupling. This provides another argument for extrapolation sampling. Once a control network is trained, not only can it be used for the free energy at different $N$ but it can also be applied to thermal expectation values across $N$ without having to train again. 

\section{Discussion}\label{sec:Discussion}

In this work we present a reinforcement learning approach for the computation of a class of path integrals that describe quantum systems at finite temperatures. The strength of this approach is that it can be used to obtain either a variational or exact result within a single framework, which is divided in two separate steps here. However, it should also be possible to combine both steps into a hybrid approach with a smooth transition from variational to exact results by, for example, slowly reducing the number of samples with respect to the stochastic averages. We also demonstrate proof of concept beyond toy models or single-particle systems and use this approach to study a medium-sized quantum rotor system with up to 25 continuous degrees of freedom. 

One of the main results of this work is that ML might be particularly suitable in this formulation because an appropriate architectural design allows you to naturally extend controls to larger systems. This suggests that applications to truly large systems in the thermodynamic limit would mainly be determined by the inference cost of the trained sampler across paths rather than by the training phase. This type of extrapolation sampling is formulated here in terms of system size, but one interesting avenue towards the future is to also explore how to extrapolate across coupling strengths or even potentials. 

Another domain where this approach is expected to generate interest is in the context of polaron physics, where the path integral approach is known to provide a remarkably fruitful description even when using very simple variational actions \cite{Feynman1955}. However, for polarons in ultracold atomic gases, the expressiveness of the trial actions that can still be used in the typical semi-analytic variational Jensen-Feynman framework becomes one of the main limiting factors to this approach \cite{Ichmoukhamedov2022}. If the optimal control formulation can be extended to the polaron setting, this would provide a major improvement over any other existing path integral variational approach for polarons in the literature. 

It is important to highlight the limitations and possible improvements of our work. The main limitation of the applicability of this approach in its current form is that for many particles at finite temperatures it can only be applied to systems where explicit symmetrization of the wavefunction is not required. To extend it to bosons or fermions in free space, path integral permutations need to be taken into account, which will be the topic of future work. To sample the free energy we introduced an additional importance sampling distribution $P$ for which a simple Gaussian approximation was sufficient for the systems explored here. However, for more complex systems it might prove necessary to integrate a learnable $P$ in the optimization scheme. Finally, we have implemented the control optimization with a methodologically simple but rather computationally expensive backpropagation through an SDE approach, which should probably be replaced by more efficient adjoint methods for further scaling. 

\section{Limitations and ethical considerations}
This work is theoretical and computational, with no involvement of human participants or personally identifiable data. Therefore, no direct ethical concerns related to subject participation or data privacy arise. Scientific and technical limitations are discussed in Sec.~\ref{sec:Discussion}.

\begin{acks}
TI is grateful for fruitful discussions with Jacques Tempere.
This project has received funding from the European Union’s Horizon Europe research and innovation programme under the Marie Skłodowska-Curie Grant agreement No 101204572 (TI). The Flatiron Institute is a division of the Simons Foundation. DS was supported by AFOSR under Award No. FA9550-21-1-0236.
\end{acks}

\balance
\bibliographystyle{ACM-Reference-Format}
\bibliography{references}

\end{document}